\documentclass[aps,prb,amsmath,amssymb,%preprint,
reprint,superscriptaddress,
longbibliography,
]{revtex4-1}

\usepackage[version=3]{mhchem} % Formula subscripts using \ce{}
\usepackage{graphicx}
\usepackage{amsmath,amssymb}
\usepackage{natbib,hyperref}% add hypertext capabilities
\hypersetup{colorlinks=true,allcolors=blue}
\usepackage{float}

\begin{document}
\author{Yanhao Tang}
\author{Wei Xie}
\affiliation
{Department of Physics and Astronomy, Michigan State University, East Lansing, Michigan 48824}
\author{Krishna C. Mandal}
\affiliation{Department of Electrical Engineering, University of South Carolina, Columbia, SC 29208, USA}
\author{John A. McGuire}
\author{Chih-Wei Lai}
\email{cwlai@msu.edu}
\affiliation
{Department of Physics and Astronomy, Michigan State University, East Lansing, Michigan 48824}

\title[]
{Linearly polarized remote-edge luminescence in GaSe nanoslabs}

\begin{abstract}
We report highly linearly polarized remote luminescence that emerges at the cleaved edges of nanoscale gallium selenide slabs tens of micrometers away from the optical excitation spot. The remote-edge luminescence (REL) measured in the reflection geometry has a degree of linear polarization above 0.90, with polarization orientation pointing toward the photoexcitation spot. The REL is dominated by an index-guided optical mode that is linearly polarized along the crystalline $c$-axis. This luminescence is from out-of-plane dipoles that are converted from in-plane dipoles through a spin-flip process at the excitation spot.
\end{abstract}

\maketitle


\begin{figure*}[htb!]\includegraphics[width = 0.9 \textwidth]{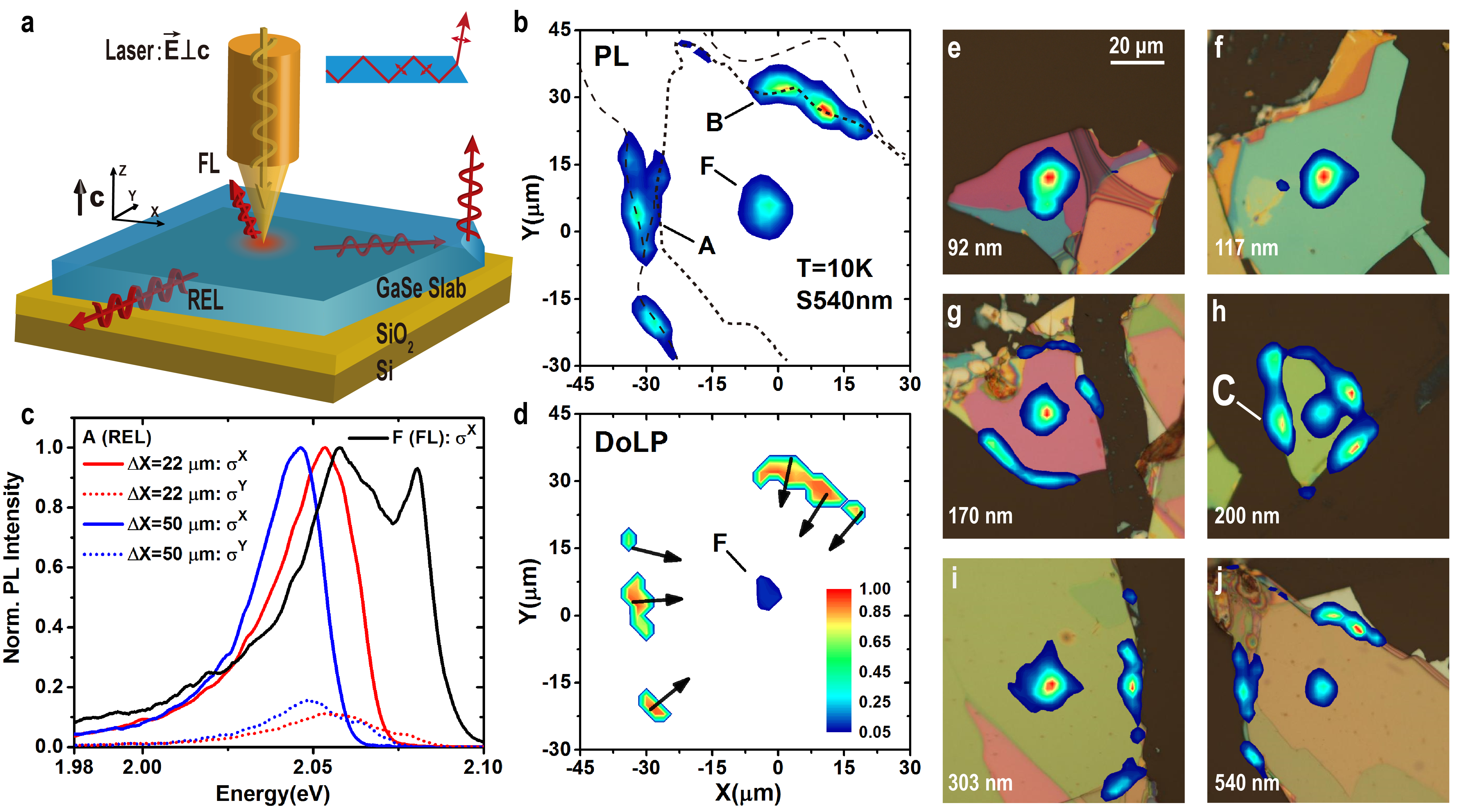}
\centering
\caption{\label{fig:plimage}(a) Schematic of the optical set-up. (b) Photoluminescence (PL) image of a 540 nm-thick GaSe slab at T = 10 K. The sample is optically pumped at the focal location F under a pump flux $P = 0.1 \ P_0$, where $P_0 = 2.6 \times 10^{14}$ cm$^{-2}$ per pulse. At $P_0$, the photoexcited carrier density is about  $3.4\times 10^{17}$ cm$^{-3}$ ($2.7\times 10^{10}$ cm$^{-2}$ per layer). The dashed lines represent the cleaved edges of the sample. Remote-edge luminescence (REL) emerges at the cleaved edges tens of micrometers away from the photoexcited spot. (c) Polarized spectra of the FL (black curve) from focal point F and the REL (red and blue curves) from location A with varying distances ($\Delta X$) from the photoexcited spot for $P$ = 0.15 $P_0$ at T = 10 K. The spectra are normalized with respect to the X-polarized ($\sigma^X$) component. Both $\sigma^X$ (TM mode, solid lines) and $\sigma^Y$ (TE mode, dashed lines) spectral components red-shift with respect to the FL with increasing $\Delta X$ as a result of anisotropic reabsorption. The anisotropic reabsorption also results in a decreasing spectrally integrated $DoLP$ with increasing $\Delta X$. The two peaks in the FL spectrum are attributed to free excitons ($\sim$2.08 eV) and exciton-carrier scattering ($\sim$ 2.06 eV), respectively. The FL is dominated by free excitons for $P \gtrsim$ 0.6 $P_0$ (not shown).  (d) Spatially resolved $DoLP$ for $P = 0.1 \ P_0$. The false color and arrows represent the $DoLP$ and polarization orientation, respectively. The measured maximal $DoLP$ is about 0.93. Note that the REL at location A (Fig.~1c) is X-polarized, whereas the REL at location B is nearly Y-polarized. The REL has a polarization oriented toward the photoexcited spot. (e--j) Overlay of optical and luminescence images of samples with thickness $d_L =$ 92, 117, 170, 200 (S200nm), 303, and 540 (S540nm). Luminescence images are measured at T = 10 K. The false colors in the luminescence images represent the relative intensities. The colors in the optical images are due to optical interference. The remote-edge luminescence (REL) is visible only for $d_L \gtrsim$ 160 nm. The central elliptical luminescence spot is the focal photoexcited region. The REL is most prevalent in the 200 nm-thick sample in which single guided TE and TM modes are present.}
\end{figure*}

Two-dimensional layered materials present a new platform for integration of optoelectronic devices and nanoscale lasers \cite{geim2013,eda2013,miro2014,fiori2014}. In this study, we exploit the anisotropic optical properties \cite{schluter1973,bourdon1971,mercier1973,le-toullec1977,piccioli1977,le-toullec1980} and unique optical selection rules \cite{gamarts1977,ivchenko1977,tang2014} in layered gallium selenide (GaSe) to generate highly linearly polarized luminescence at the remote cleaved edges of GaSe slabs of about 200 layers or more (Fig.~\ref{fig:plimage}a). The in-plane dipoles ($d_\perp$) generated at the photoexcited spot are converted to out-of-plane dipoles ($d_\parallel$) through a spin-flip process. Subsequent luminescence from $d_\parallel$ propagates with wave vector orthogonal to the incident optical excitation beam through an index-guided optical mode that is linearly polarized along the crystalline $c$-axis. Remote luminescence emerges at the cleaved edges in the reflection geometry as a result of backscattering. This remote-edge luminescence (REL) has a degree of linear polarization above 0.90, with polarization orientation pointing toward the photoexcitation spot independent of the photoexcitation polarization. In contrast, the luminescence at the focal photoexcited spot (FL) is unpolarized under a linearly polarized optical excitation.

%The unique REL distribution and polarization patterns in GaSe nanoslabs were not previously reported in layered materials.
In most bulk semiconductors, the optical orientation and alignment of electron-hole (\emph{e-h}) pairs and consequent polarized luminescence are limited \cite{meier1984}. In quasi-2D systems, such as semiconductor quantum wells, photoluminescence is typically unpolarized or circularly polarized, depending on the photoexcitation energy and polarization as well as the band structure. Highly linearly polarized luminescence or lasing has been observed in quasi-1D colloidal nanowires \cite{hu2001,wang2001b,johnson2003} with strong dielectric confinement, as well as in quasi-0D quantum dots with a sizable anisotropy induced by strain or electron-hole exchange interactions \cite{takagahara1993,gammon1996,dzhioev1998,paillard2001}. Recently, in 2D atomically thin transition metal dichalcogenides (TMDs), polarized luminescence has been shown to result from intra- and interlayer excitations \cite{schuller2013} or valley coherence \cite{jones2013,jones2014}. In 2D TMDs, though, the luminescence yield only becomes significant at the monolayer level when a crossover from indirect to direct gap occurs.

GaSe is a layered semiconductor characterized by covalently bound gallium and selenium atoms within a single layer and weak van der Waals-type interlayer interactions \cite{fernelius1994}. The layers are stacked along the crystallographic $c$-axis and form several polytypes with different stacking orders. The noncentrosymmetric $\epsilon$-GaSe is the most widely studied monochalcogenide among the group--III monochalcogenides because of its high optical second-order nonlinearity \cite{abdullaev1972,fernelius1994,allakhverdiev2009}. The \emph{layered} structure of $\epsilon$-GaSe results in highly anisotropic optical properties, as demonstrated by the distinct absorbance for light with $\vec{E} \perp c$ and $\vec{E} \parallel c$ \cite{schluter1973,bourdon1971,mercier1973,le-toullec1977,piccioli1977,le-toullec1980}. Polarized spontaneous and stimulated emissions also occur near the quasi-direct gap ($E_g\sim$2 eV) under optical excitation/detection at oblique angles \cite{nahory1971,ugmori1973,voitchovsky1974,mercier1975,moriya1976,moriya1976a,bernier1986,pavesi1989,capozzi1993}. Nanoscale slabs of similar polar (pyroelectric) group-III monochalcogenide semiconductors have high optical nonlinearities and allow direct optical access for control of light--matter interactions and polarization properties, as well as index-guiding of light.
 
%\paragraph{Results.}
 
In this study, we investigate the spectral, \emph{polarization}, and \emph{dynamic} characteristics of the luminescence emanating at cleaved edges of GaSe slabs with a thickness $d_L$ of 160 nm or more. The GaSe nanoslabs are mechanically exfoliated from a bulk $\epsilon$-GaSe crystal \cite{mandal2008a} and deposited onto a \ce{Si} substrate with a 90 nm \ce{SiO2} layer. These GaSe dielectric slabs on \ce{SiO2}/\ce{Si} form the simplest optical waveguides (Materials and Methods) \cite{marcuse1991}. The thinnest GaSe that can sustain one guided transverse electric (TE) mode ($\vec{E} \perp c$) and transverse magnetic (TM) mode ($\vec{E} \parallel c$) for luminescence near the band edge (wavelength $\lambda \approx$ 600 nm) is about 160 nm thick. The normally incident 2 ps optical excitation pulses create in-plane dipoles ($d_\perp$) that are rapidly converted to out-of-plane dipoles ($d_\parallel$) through spin-flip of electrons or holes (Fig.~1a) \cite{gamarts1977,ivchenko1977,tang2014}. The index-guided radiation from $d_\parallel$ at the photoexcited spot propagates to the cleaved edges of the GaSe slabs. Under pulsed photoexcitation, the guided TM mode of $d_\parallel$ luminescence dominates over the TE mode of $d_\perp$ luminescence because the radiative recombination rate of $d_\parallel$ is about 30 times higher than that of $d_\perp$. As a result, highly linearly polarized TM mode luminescence emerges in the reflection geometry at cleaved edges when backscattering is allowed in the presence of irregularities. We refer to such polarized remote luminescence at cleaved edges as remote-edge luminescence (REL). The REL typically has a degree of linear polarization $DoLP >$ 0.90. Note that enhanced photoluminescence (PL) has also been observed at edges of \ce{WS2} platelets where PL emanates from the \emph{photoexcited spot} \cite{gutierrez2012}. In the experiments presented in this study, the in-plane ($d_\perp$) and out-of-plane ($d_\parallel$) dipoles are within the same layer, and the REL occurs tens of micrometers away from the photoexcited spot. 

In Fig.~\ref{fig:plimage}, we study the spectral and polarization characteristics of the REL from a 540 nm-thick GaSe slab under excitation energy $E_p$ = 2.138 eV at T = 10 K. Intense luminescence emerges at the cleaved edges of the GaSe slab even when the optical excitation spot is tens of micrometers away (Fig.~\ref{fig:plimage}b). Spectrally, the REL shows an apparent red shift compared with the luminescence at the focal photoexcited spot (FL). Additionally, the REL is highly linearly polarized ($DoCP \rightarrow$ 0.93) , whereas the FL is unpolarized under linearly polarized photoexcitation (Fig.~\ref{fig:plimage}d). Considering the REL as an index-guided mode of luminescence originating from the photoexcited spot, one can attribute the spectral redshift to anisotropic reabsorption \cite{moriya1976a} when luminescence propagates in-plane for a distance of tens of micrometers (Fig.~1c). However, for a 540 nm-thick GaSe dielectric slab on \ce{SiO2}, about four guided TE modes and TM modes are allowed. In the presence of both TE and TM modes, the REL is expected to be unpolarized when the orientations of the photoexcited dipoles at the focal spot are random and isotropic. On the contrary, the measured luminescence polarization image (Fig.~\ref{fig:plimage}d) indicates that the polarization orientations of the REL point toward the photoexcited spot. Moreover, depending on the location and sample thickness, the intensity (emission flux) of the REL can exceed that of the FL (Fig.~\ref{fig:plimage}f--j). Such a polarization orientation pattern and intensity distribution suggest that the REL originates from out-of-plane dipoles ($d_\parallel$), with the electric field vector parallel to the crystalline $c$-axis ($\vec{E} \parallel c$).   

%figure 2
\begin{figure}[htb!]\includegraphics[width = 0.3 \textwidth]{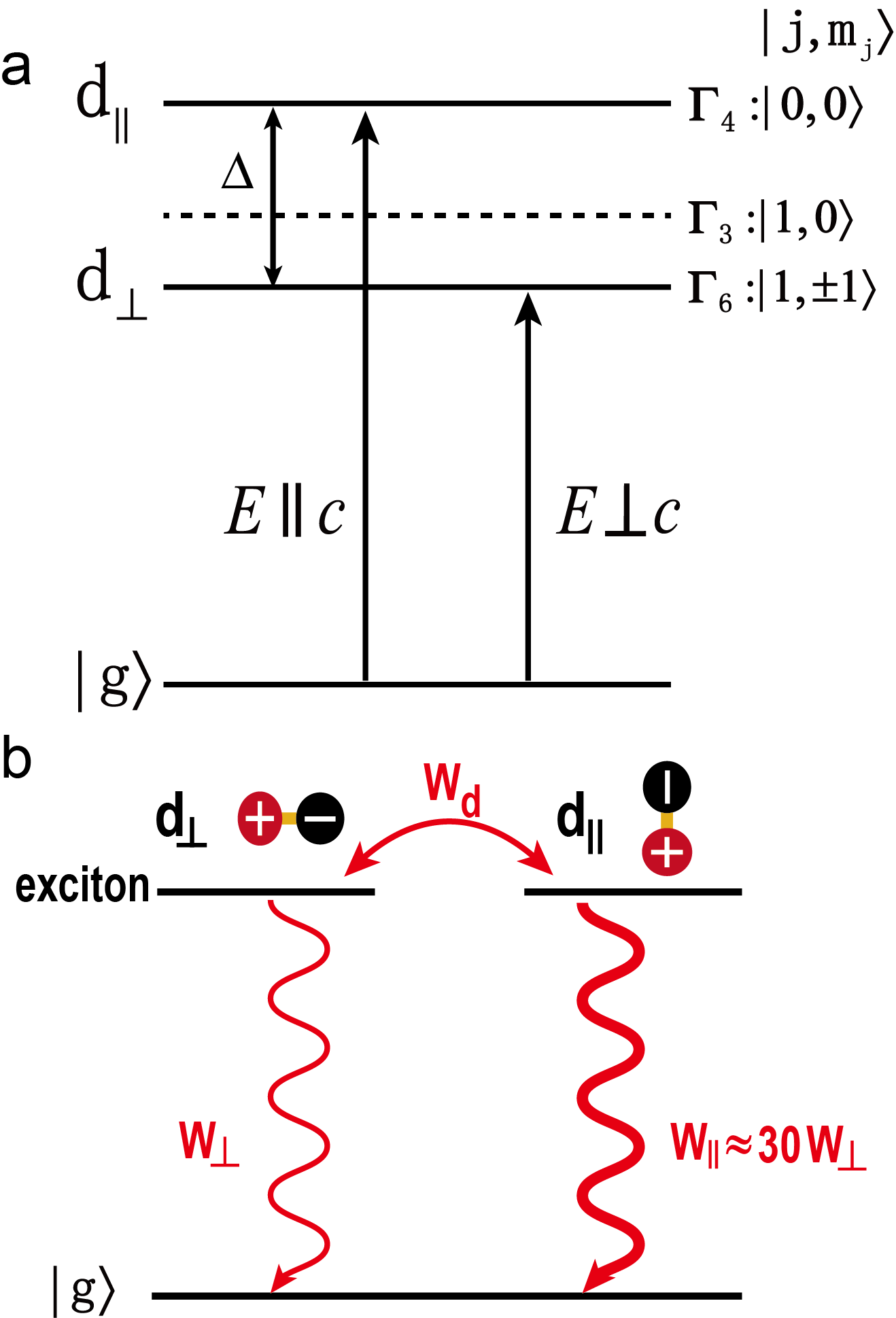}
	\caption{ \label{fig:selection}(a) Schematic exciton levels in $\epsilon$-GaSe at the $\Gamma$ point and the representations to which the states at the $\Gamma$-point belong when including spin--orbit interaction. The energy splitting $\Delta$ between the $\Gamma_4$ ($d_\parallel$) and $\Gamma_6$ ($d_\perp$) exciton states is approximately 2 meV \cite{gamarts1977,ivchenko1977}. (b) Schematic of the rate equation model. 
	}
\end{figure}

%\subsection*{Optical selection rules} 
To understand the unique polarization of the REL, we need to consider the selection rules and anisotropic optical constants in GaSe. We first examine the optical selection rules in GaSe and illustrate the spin-flip-induced conversion between the in-plane and out-of-plane dipoles (Fig.~\ref{fig:selection}). The Se 4$p_z$ states lie 1.2 and 1.6 eV above the Se 4$p_{x,y}$ states as a result of the crystal field and spin--orbit interaction. Near the $\Gamma$ point, direct optical transitions between the $p_z$-like uppermost valence band and $s$-like lowermost conduction band are dipole-allowed for light with the electric field parallel to the $c$-axis ($\vec{E} \parallel c$). For light with $\vec{E} \perp c$, the transitions become weakly allowed due to spin--orbit interaction \cite{mercier1973}. The spin-dependent optical selection rules can be best understood in the two-particle (exciton) representation (Fig.~\ref{fig:selection}a) \cite{gamarts1977,ivchenko1977}. Optical excitation with $\vec{E} \perp c$ creates excitons with $\Gamma_6$ symmetry ($d_\perp$ dipoles), which is analogous to a triplet state ($\uparrow \Uparrow$ and $\downarrow \Downarrow$). On the other hand, optical excitation with $\vec{E} \parallel c$ results in excitons with $\Gamma_4$ symmetry ($d_\parallel$ dipoles), which is analogous to a singlet state ($\uparrow \Downarrow - \downarrow \Uparrow$). Therefore, a spin flip of either electron or hole of the exciton (\emph{e-h} pair) results in the conversion between $d_\perp$ and $d_\parallel$ (Fig.~\ref{fig:selection}b).

Experimentally, the absorption coefficient near the band edge for $\vec{E} \perp c$ is about $3 \times 10^{-3}$ cm$^{-1}$, which is a factor of 1/30 of that for light with $\vec{E} \parallel c$. Assuming that the spontaneous radiative recombination rate is proportional to the absorption coefficient, we approximate the relative radiative recombination rates of the in-plane and out-plane dipoles to be $W^r_\parallel/W^r_\perp \approx$ 30. As a result, under ps pulse excitation, luminescence is dominated by the radiative recombination of $d_\parallel$ when the conversion between $d_\perp$ and $d_\parallel$ is fast compared with the recombination rates. The linear polarization of the REL is approximately $(W_\parallel^r-W_\perp^r)/(W_\parallel^r+W_\perp^r) \approx$ 0.94, which is consistent with the measured $DoLP$.

%figure 3
\begin{figure}[htb!]\includegraphics[width = 0.4 \textwidth]{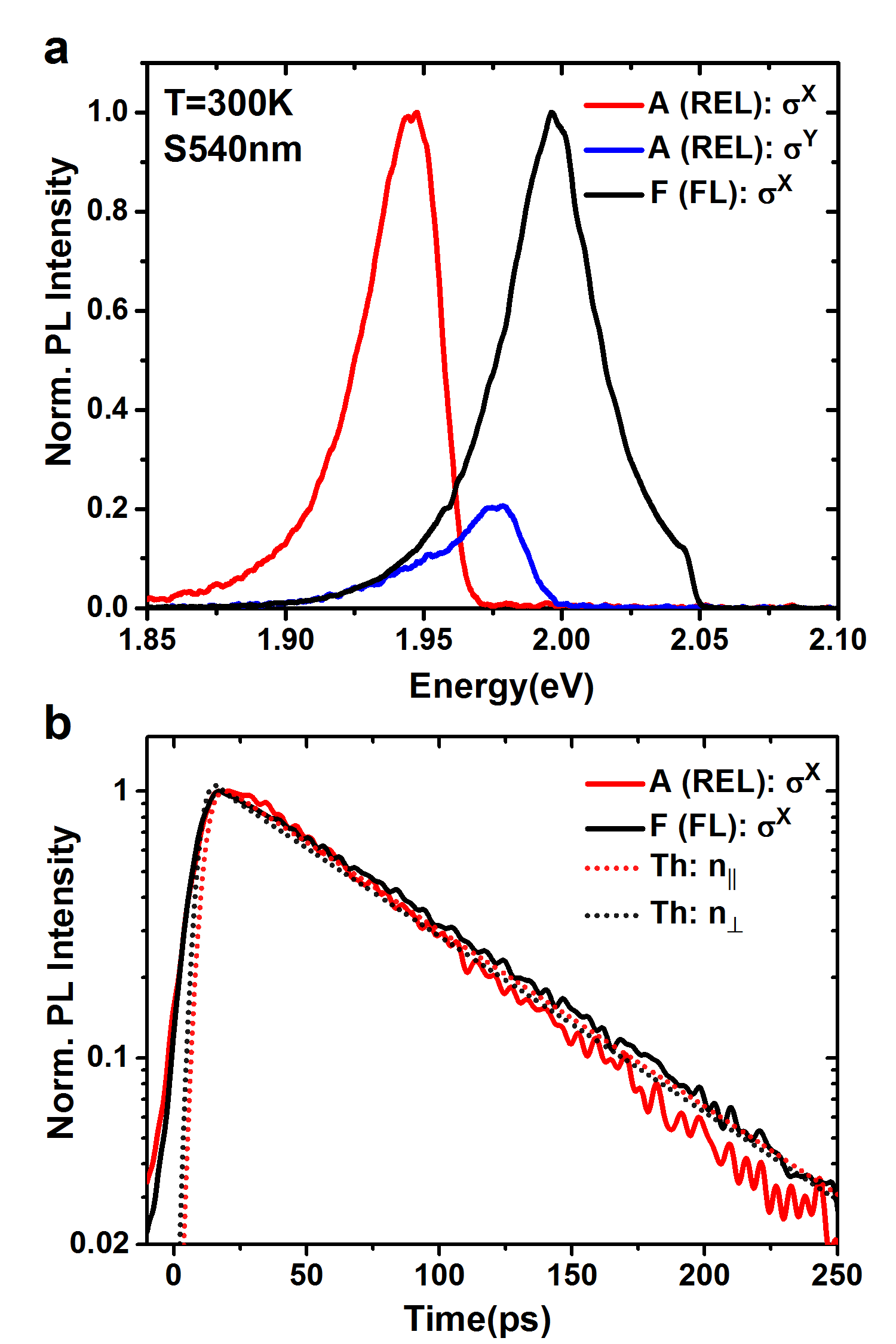}
	\caption{ \label{fig:rtpl}(a) Polarized spectra of the FL (black curve) and the REL from location A (red and blue curves) in a 540 nm-thick GaSe slab at room temperature. The pump energy $E_p =$ 2.087 eV and the pump flux $P$ = 1.0 $P_0$. At room temperature, the FL is largely due to free excitons, and the backscattered REL intensity is considerably weaker than the FL (about a factor of four here). The $DoLP$ is above 0.93 for $E <$ 1.95 eV, where the anisotropic reabsorption for $\sigma^X$ (TM mode) and $\sigma^Y$ (TE mode) is negligible. (b) Time-dependent FL and REL. The FL (solid black curve) and the REL (solid red curve) show similar rise and decay times. The dotted black (red) curve is the calculated $n_\perp$ ($n_\parallel$), which corresponds to the time-dependent luminescence from $d_\perp$ ($d_\parallel$) dipoles. The theoretical curves are calculated with a rate equation model and convolved with an 8 ps (FWHM) instrument response. The fitting parameters are as follows: $W_\perp^r$ = 1/1000, $W_\parallel^r$ = 30 $W_\perp^r$, $W_d$ = 1/3 [ps$^{-1}$].
	}
\end{figure}

In Fig.~\ref{fig:rtpl}, we study the spectral and dynamic characteristics of the REL at room temperature. The REL remains highly linearly polarized, with a polarization orientation pointing to the photoexcited spot. The FL and REL are nearly synchronized temporally without an identifiable time delay in the REL. Therefore, the REL cannot originate from carriers that are photoexcited at the focal spot and transported to the cleaved edges. We attribute the REL to the back scattered light from an index-guided optical TM mode with $\vec{E} \parallel c$. The $d_\parallel$  converted from the photoexcited $d_\perp$ radiate into this TM mode, which propagates in-plane from the photoexcited spot to the cleaved edges at the speed of light in the GaSe slab. The sub-10-ps rise time of the REL in Fig.~\ref{fig:rtpl} indicates that energy relaxation and conversion of photoexcited $d_\perp$ and $d_\parallel$ are less than 10 ps at room temperature \cite{gamarts1977,ivchenko1977,tang2014}.

%figure 4
\begin{figure}[htbp]\includegraphics[width = 0.45 \textwidth]{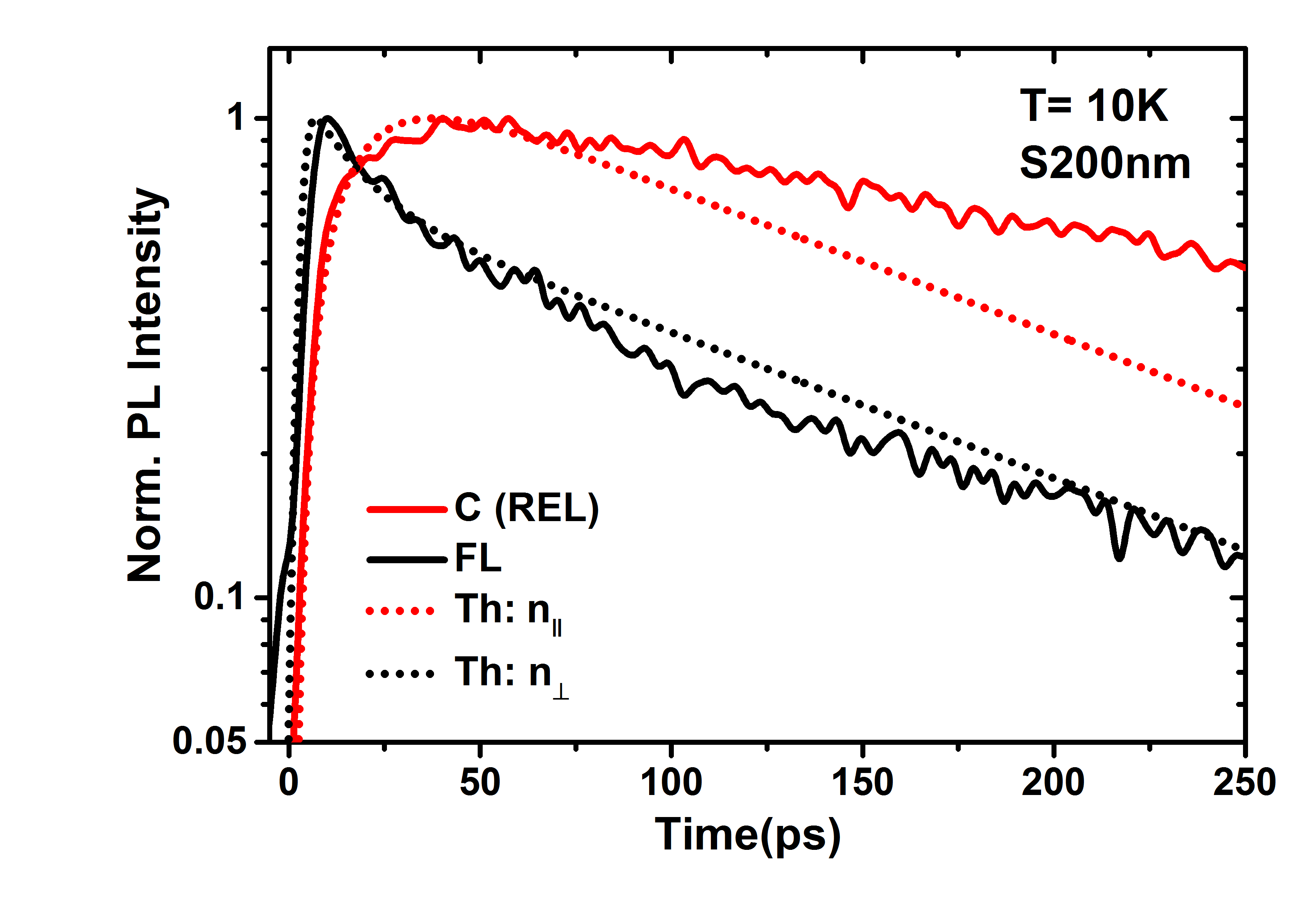}
\caption{ \label{fig:dynamics} Time-dependent FL (solid black line) and REL (solid red line) in a 200 nm-thick sample (S200nm). The REL is measured at location C, as indicated in Fig.~1h. The dotted black (red) line is the calculated time-dependent $n_\perp$ ($n_\parallel$) convolved with a 4 ps (FWHM) instrument response function. The fitting parameters are as follows: $W_\perp^r$ = 1/2000, $W_\parallel^r$ = 30 $W_\perp^r$, $W_d$ = 1/30 [ps$^{-1}$]. The slower decay of the REL relative to the FL is likely due to reabsorption, which results in the dominance of the REL by the low energy portion of the exciton luminescence. 
%The REL is dominated by the low-energy spectral part of the exciton luminescence as a result of reabsorption and appears to have a slightly longer decay than the FL. 
}	
\end{figure}

The spin-flip-induced conversion between the in-plane and out-of-plane dipoles is evident in the time-dependent PL measurements at cryogenic temperature (T = 10 K), where the spin-flip rate is significantly reduced (Fig.~\ref{fig:dynamics}). The FL arises within 10 ps after the pulse excitation, whereas the REL reaches its maximum about 50 ps afterward. The nearly instantaneous rise of the FL is the result of a sub-5-ps energy/momentum relaxation of photoexcited carriers, whereas the considerable delay of $\sim$30--50 ps rise time in the REL is due to the $\sim$ 30 ps spin-flip time constant at T = 10 K. 

%\subsection*{Modeling}
The polarized time-dependent luminescence is reproduced by a rate equation model. Considering a spin-flip process and anisotropic radiative recombination rates for the in-plane and out-of-plane dipoles, we describe the time-dependent FL and REL with the following set of coupled differential equations:
\begin{align*}
\dot{n}_\perp (t) &=  - W^r_\perp n_\perp(t) - {W_d} \left[ {n_\perp(t) - n_\parallel(t)} \right] + G(t) P \ , \\ 
\dot{n}_\parallel (t) &=  - W^r_\parallel n_\parallel(t) + {W_d} \left[ {n_\perp(t) - n_\parallel(t)} \right].
\end{align*}

\noindent The optical generation rate is approximated by a 2 ps Gaussian pulse $G(t) P$, where $P$ is the pump flux. $n_\perp$ and $n_\parallel$ are the populations of the in-plane ($d_\perp$) and out-of-plane ($d_\parallel$) dipoles, whereas $W^r_\perp$ and $W^r_\parallel$ are the corresponding radiative recombination rates. $W_d$ is the spin-flip rate that leads to the conversion between $d_\perp$ and $d_\parallel$. In this simple model, we neglect the energy relaxation of nonresonantly excited carriers and non-radiative recombination loss. The time-dependent REL agrees with the calculated temporal evolution of $n_\parallel$, and indicates a rise time of $\sim$30 ps that is consistent with the previously measured spin-flip time constant $\tau_s \approx 1/W_d$ \cite{tang2014}. Similarly, the time-dependent FL is associated with the temporal evolution of $n_\perp$. 

In summary, we have observed highly linearly polarized remote-edge luminescence in GaSe platelets with a thickness of 160 nm or more. The REL is due to the guided TM mode of luminescence from the out-of-plane dipoles that are converted from the in-plane dipoles photoexcited by normally incident ps pulses at a remote spot. One can anticipate linearly polarized stimulated emission or lasing in a GaSe platelets with a thickness of $\sim$ 200-300 nm and a lateral size of several micrometers when optical cavity effects are enhanced. 

\paragraph*{Materials and Methods.}
\emph{Sample preparation.} GaSe nanoslabs are mechanically exfoliated from a Bridgman grown \ce{GaSe} crystal \cite{mandal2008a} and deposited onto a 90-nm \ce{SiO2}/\ce{Si} substrate. The samples studied in this study are mainly the $\epsilon$ modifications. A $\epsilon$-GaSe crystal has ABA stacking of the individual layers and belongs to space group $D_{3h}^1/P\bar{6}m2$ (\#187) and point group $D_{3h}$. An individual layer consists of four planes of Se-Ga-Ga-Se, with the Ga-Ga bond normal to the layer plane arranged on a hexagonal lattice and the Se anions located in the eclipsed conformation when viewed along the $c$-axis. The band-edge luminescence and absorption in $\epsilon$-GaSe are around 630 nm (2.0 eV) and 595 nm (2.1 eV) at room temperature and T = 10 K, respectively. Samples are attached to a copper cold finger in an optical liquid-helium-flow cryostat and kept in a vacuum for all optical measurements.
   
\emph{Optical set-ups.} A 2 ps pulse laser beam is focused through an objective (N.A. = 0.28) to an area of about 80 $\mu$m$^2$. As a result, the electric field vector of the pump laser is orthogonal to the crystalline $c$-axis ($\vec{E} \perp c$). The photoexcitation density is from $\approx 2\times10^{16}$ cm$^{-3}$ to $3.4\times10^{17}$ cm$^{-3}$ ($2.7\times10^{-10}$ cm$^{-2}$ per layer) considering the absorption coefficient at 2.1 eV ($\approx10^3$ cm$^{-1}$ for $\vec{E} \perp c$) and Fresnel loss from reflection. The photoexcited density is below the Mott transition density in GaSe \cite{pavesi1989,capozzi1993}, which is about $4\times10^{17}$ cm$^{-3}$. The PL is collected through the same objective in the reflection geometry. The time-integrated PL spectra are measured with an imaging spectrometer (focal length: 750 mm; grating: 300 grooves/mm) equipped with a liquid-nitrogen cooled CCD camera. The time-resolved PL measurements are obtained with a streak camera system with a $\sim$4--10 ps (FWHM) temporal resolution depending on the exposure time. The polarization properties of the pump laser and PL are controlled/analyzed by a combination of polarizers and liquid-crystal devices without mechanical moving parts.

\emph{Polarization.} Linearly polarized light with horizontal (vertical) polarization is defined as $\sigma^X$ ($\sigma^Y$). The circularly polarized pump or luminescence with angular momentum $+\hbar$ ($-\hbar$) along the pump laser wavevector $\hat{k} \parallel \hat{z}$ is defined as $\sigma^{+}$ ($\sigma^{-}$). The polarization state is characterized by the Stokes vector $\{S_0, S_1, S_2, S_3\}$. $S_0$ is the flux and is determined as $S_0 = I^{X}+I^{Y}$. The Stokes vector can be normalized by its flux $S_0$ to the Stokes three-vector $s = \{s_1, s_2, s_3\}$. $s_1 = (I^{X}-I^{Y})/S_0$, $s_2 = (I^{45^\circ}-I^{135^\circ})/S_0$, and $s_3 = (I^{+}-I^{-})/S_0$. $I^{+}$, $I^{-}$, $I^{X}$, $I^{Y}$, $I^{45^\circ}$, and $I^{135^\circ}$ are measured intensities of the circularly or linearly polarized components. The degree of linear polarization is represented by $DoLP \equiv \sqrt{s_1^2+s_2^2}$, and the polarization orientation angle is determined by $\tan^{-1}(s_2/s_1)$. The accuracy of the measurements of the $DoLP$ are within 1-2\%.

\emph{Guided TE and TM modes \cite{marcuse1991}.} We define a parameter $V = (n_1^2-n_2^2)^{1/2} \, k \, d_L$, which includes the difference in the squares of the refractive indices of the core (i.e., $n_1$ of the GaSe slab) and the substrate (i.e., $n_2$ of the \ce{SiO2} layer), the vacuum wavelength ($\lambda_0$) and wavenumber ($2 \pi/\lambda_0$), and the thickness of the core slab ($d_L$). The total number of TE or TM modes is then determined by the following equation:
\begin{equation}
	N = \frac{1}{\pi} \left[ V - \arctan \left(\eta \sqrt{\frac{n_2^2-n_3^2} {n_1^2-n_2^2}} \right) \right]_{int} \nonumber.
\end{equation}

\noindent where $n_3$ is the refractive index of the capping layer (here vacuum), and the symbol $[\,]_{int}$ indicates that N is the integer part of the number in brackets. The parameter $\eta$ is defined as 
\begin{equation}
	\eta = 
	\begin{cases}
	1, \, \text{for TE modes}. \\
	n_1^2/n_3^2, \, \text{for TM modes} \nonumber.
	\end{cases}
\end{equation}

The number of TE and TM modes is usually the same for most $d_L$. However, the number of TE modes can exceed that of the number of TM modes by one for specific ranges of $d_L$ because $n_1/n_3 > 1$. For example, two TE modes and one TM modes exist for 243 nm $\lesssim d_L \lesssim$ 276 nm. Using $n_1 =  3.00$ (\ce{GaSe}), $n_2 = 1.46$ (\ce{SiO2}), $n_3 = 1.00$ (vacuum), and $\lambda_0$ = 600 nm (band-edge exciton luminescence at T$\sim$10 K), we determine the thickness required to sustain only one guided TM mode and one TE mode to be about 162 to 242 nm.

\begin{acknowledgements}
This work was supported by NSF grant DMR-09055944 as well as a start-up funding and the Cowen endowment at Michigan State University. This research has used the W. M. Keck Microfabrication Facility. We thank Norman Birge, Brage Golding, and Bhanu Mahanti for the discussions.
\end{acknowledgements}



\begin{thebibliography}{39}%
\makeatletter
\providecommand \@ifxundefined [1]{%
 \@ifx{#1\undefined}
}%
\providecommand \@ifnum [1]{%
 \ifnum #1\expandafter \@firstoftwo
 \else \expandafter \@secondoftwo
 \fi
}%
\providecommand \@ifx [1]{%
 \ifx #1\expandafter \@firstoftwo
 \else \expandafter \@secondoftwo
 \fi
}%
\providecommand \natexlab [1]{#1}%
\providecommand \enquote  [1]{``#1''}%
\providecommand \bibnamefont  [1]{#1}%
\providecommand \bibfnamefont [1]{#1}%
\providecommand \citenamefont [1]{#1}%
\providecommand \href@noop [0]{\@secondoftwo}%
\providecommand \href [0]{\begingroup \@sanitize@url \@href}%
\providecommand \@href[1]{\@@startlink{#1}\@@href}%
\providecommand \@@href[1]{\endgroup#1\@@endlink}%
\providecommand \@sanitize@url [0]{\catcode `\\12\catcode `\$12\catcode
  `\&12\catcode `\#12\catcode `\^12\catcode `\_12\catcode `\%12\relax}%
\providecommand \@@startlink[1]{}%
\providecommand \@@endlink[0]{}%
\providecommand \url  [0]{\begingroup\@sanitize@url \@url }%
\providecommand \@url [1]{\endgroup\@href {#1}{\urlprefix }}%
\providecommand \urlprefix  [0]{URL }%
\providecommand \Eprint [0]{\href }%
\providecommand \doibase [0]{http://dx.doi.org/}%
\providecommand \selectlanguage [0]{\@gobble}%
\providecommand \bibinfo  [0]{\@secondoftwo}%
\providecommand \bibfield  [0]{\@secondoftwo}%
\providecommand \translation [1]{[#1]}%
\providecommand \BibitemOpen [0]{}%
\providecommand \bibitemStop [0]{}%
\providecommand \bibitemNoStop [0]{.\EOS\space}%
\providecommand \EOS [0]{\spacefactor3000\relax}%
\providecommand \BibitemShut  [1]{\csname bibitem#1\endcsname}%
\let\auto@bib@innerbib\@empty
%</preamble>
\bibitem [{\citenamefont {Geim}\ and\ \citenamefont
  {Grigorieva}(2013)}]{geim2013}%
  \BibitemOpen
  \bibfield  {author} {\bibinfo {author} {\bibfnamefont {A.~K.}\ \bibnamefont
  {Geim}}\ and\ \bibinfo {author} {\bibfnamefont {I.~V.}\ \bibnamefont
  {Grigorieva}},\ }\bibfield  {title} {\enquote {\bibinfo {title} {{Van der
  Waals} heterostructures},}\ }\href {\doibase 10.1038/nature12385} {\bibfield
  {journal} {\bibinfo  {journal} {Nature}\ }\textbf {\bibinfo {volume} {499}},\
  \bibinfo {pages} {419--425} (\bibinfo {year} {2013})}\BibitemShut {NoStop}%
\bibitem [{\citenamefont {Eda}\ and\ \citenamefont {Maier}(2013)}]{eda2013}%
  \BibitemOpen
  \bibfield  {author} {\bibinfo {author} {\bibfnamefont {G.}~\bibnamefont
  {Eda}}\ and\ \bibinfo {author} {\bibfnamefont {S.~A.}\ \bibnamefont
  {Maier}},\ }\bibfield  {title} {\enquote {\bibinfo {title} {Two-dimensional
  crystals: Managing light for optoelectronics.}}\ }\href {\doibase
  10.1021/nn403159y} {\bibfield  {journal} {\bibinfo  {journal} {ACS Nano}\
  }\textbf {\bibinfo {volume} {7}},\ \bibinfo {pages} {5660--5665} (\bibinfo
  {year} {2013})}\BibitemShut {NoStop}%
\bibitem [{\citenamefont {Mir{\'o}}\ \emph {et~al.}(2014)\citenamefont
  {Mir{\'o}}, \citenamefont {Audiffred},\ and\ \citenamefont
  {Heine}}]{miro2014}%
  \BibitemOpen
  \bibfield  {author} {\bibinfo {author} {\bibfnamefont {P.}~\bibnamefont
  {Mir{\'o}}}, \bibinfo {author} {\bibfnamefont {M.}~\bibnamefont {Audiffred}},
  \ and\ \bibinfo {author} {\bibfnamefont {T.}~\bibnamefont {Heine}},\
  }\bibfield  {title} {\enquote {\bibinfo {title} {An atlas of two-dimensional
  materials},}\ }\href {\doibase 10.1039/C4CS00102H} {\bibfield  {journal}
  {\bibinfo  {journal} {Chem. Soc. Rev.}\ }\textbf {\bibinfo {volume} {43}},\
  \bibinfo {pages} {6537--6554} (\bibinfo {year} {2014})}\BibitemShut {NoStop}%
\bibitem [{\citenamefont {Fiori}\ \emph {et~al.}(2014)\citenamefont {Fiori},
  \citenamefont {Bonaccorso}, \citenamefont {Iannaccone}, \citenamefont
  {Palacios}, \citenamefont {Neumaier}, \citenamefont {Seabaugh}, \citenamefont
  {Banerjee},\ and\ \citenamefont {Colombo}}]{fiori2014}%
  \BibitemOpen
  \bibfield  {author} {\bibinfo {author} {\bibfnamefont {G.}~\bibnamefont
  {Fiori}}, \bibinfo {author} {\bibfnamefont {F.}~\bibnamefont {Bonaccorso}},
  \bibinfo {author} {\bibfnamefont {G.}~\bibnamefont {Iannaccone}}, \bibinfo
  {author} {\bibfnamefont {T.}~\bibnamefont {Palacios}}, \bibinfo {author}
  {\bibfnamefont {D.}~\bibnamefont {Neumaier}}, \bibinfo {author}
  {\bibfnamefont {A.}~\bibnamefont {Seabaugh}}, \bibinfo {author}
  {\bibfnamefont {S.~K.}\ \bibnamefont {Banerjee}}, \ and\ \bibinfo {author}
  {\bibfnamefont {L.}~\bibnamefont {Colombo}},\ }\bibfield  {title} {\enquote
  {\bibinfo {title} {Electronics based on two-dimensional materials},}\ }\href
  {\doibase 10.1038/nnano.2014.207} {\bibfield  {journal} {\bibinfo  {journal}
  {Nature Nanotechno.}\ }\textbf {\bibinfo {volume} {9}},\ \bibinfo {pages}
  {768} (\bibinfo {year} {2014})}\BibitemShut {NoStop}%
\bibitem [{\citenamefont {Schl\"{u}ter}(1973)}]{schluter1973}%
  \BibitemOpen
  \bibfield  {author} {\bibinfo {author} {\bibfnamefont {M.}~\bibnamefont
  {Schl\"{u}ter}},\ }\bibfield  {title} {\enquote {\bibinfo {title} {The
  electronic structure of {GaSe}},}\ }\href {\doibase 10.1007/BF02726713}
  {\bibfield  {journal} {\bibinfo  {journal} {Nuovo Cimento B}\ }\textbf
  {\bibinfo {volume} {13}},\ \bibinfo {pages} {313--360} (\bibinfo {year}
  {1973})}\BibitemShut {NoStop}%
\bibitem [{\citenamefont {Bourdon}\ and\ \citenamefont
  {Khelladi}(1971)}]{bourdon1971}%
  \BibitemOpen
  \bibfield  {author} {\bibinfo {author} {\bibfnamefont {A.}~\bibnamefont
  {Bourdon}}\ and\ \bibinfo {author} {\bibfnamefont {F.}~\bibnamefont
  {Khelladi}},\ }\bibfield  {title} {\enquote {\bibinfo {title} {Selection rule
  in the fundamental direct absorption of {GaSe}},}\ }\href {\doibase
  10.1016/0038-1098(71)90301-2} {\bibfield  {journal} {\bibinfo  {journal}
  {Solid State Commun.}\ }\textbf {\bibinfo {volume} {9}},\ \bibinfo {pages}
  {1715 -- 1717} (\bibinfo {year} {1971})}\BibitemShut {NoStop}%
\bibitem [{\citenamefont {Mercier}\ \emph {et~al.}(1973)\citenamefont
  {Mercier}, \citenamefont {Mooser},\ and\ \citenamefont
  {Voitchovsky}}]{mercier1973}%
  \BibitemOpen
  \bibfield  {author} {\bibinfo {author} {\bibfnamefont {A.}~\bibnamefont
  {Mercier}}, \bibinfo {author} {\bibfnamefont {E.}~\bibnamefont {Mooser}}, \
  and\ \bibinfo {author} {\bibfnamefont {J.~P.}\ \bibnamefont {Voitchovsky}},\
  }\bibfield  {title} {\enquote {\bibinfo {title} {Near edge optical absorption
  and luminescence of {GaSe}, {GaS} and of mixed crystals},}\ }\href {\doibase
  10.1016/0022-2313(73)90070-7} {\bibfield  {journal} {\bibinfo  {journal} {J.
  Lumin.}\ }\textbf {\bibinfo {volume} {7}},\ \bibinfo {pages} {241 -- 266}
  (\bibinfo {year} {1973})}\BibitemShut {NoStop}%
\bibitem [{\citenamefont {Le~Toullec}\ \emph {et~al.}(1977)\citenamefont
  {Le~Toullec}, \citenamefont {Piccioli}, \citenamefont {Mejatty},\ and\
  \citenamefont {Balkanski}}]{le-toullec1977}%
  \BibitemOpen
  \bibfield  {author} {\bibinfo {author} {\bibfnamefont {R.}~\bibnamefont
  {Le~Toullec}}, \bibinfo {author} {\bibfnamefont {N.}~\bibnamefont
  {Piccioli}}, \bibinfo {author} {\bibfnamefont {M.}~\bibnamefont {Mejatty}}, \
  and\ \bibinfo {author} {\bibfnamefont {M.}~\bibnamefont {Balkanski}},\
  }\bibfield  {title} {\enquote {\bibinfo {title} {Optical constants of
  {$\epsilon$-GaSe}},}\ }\href {\doibase 10.1007/BF02723483} {\bibfield
  {journal} {\bibinfo  {journal} {Nuovo Cimento B}\ }\textbf {\bibinfo {volume}
  {38}},\ \bibinfo {pages} {159--167} (\bibinfo {year} {1977})}\BibitemShut
  {NoStop}%
\bibitem [{\citenamefont {Piccioli}\ \emph {et~al.}(1977)\citenamefont
  {Piccioli}, \citenamefont {Le~Toullec}, \citenamefont {Mejatty},\ and\
  \citenamefont {Balkanski}}]{piccioli1977}%
  \BibitemOpen
  \bibfield  {author} {\bibinfo {author} {\bibfnamefont {N.}~\bibnamefont
  {Piccioli}}, \bibinfo {author} {\bibfnamefont {R.}~\bibnamefont
  {Le~Toullec}}, \bibinfo {author} {\bibfnamefont {M.}~\bibnamefont {Mejatty}},
  \ and\ \bibinfo {author} {\bibfnamefont {M.}~\bibnamefont {Balkanski}},\
  }\bibfield  {title} {\enquote {\bibinfo {title} {Refractive index of {GaSe}
  between 0.45 $\mu$m and 330 $\mu$m},}\ }\href {\doibase 10.1364/AO.16.001236}
  {\bibfield  {journal} {\bibinfo  {journal} {Appl. Opt.}\ }\textbf {\bibinfo
  {volume} {16}},\ \bibinfo {pages} {1236--1238} (\bibinfo {year}
  {1977})}\BibitemShut {NoStop}%
\bibitem [{\citenamefont {Le~Toullec}\ \emph {et~al.}(1980)\citenamefont
  {Le~Toullec}, \citenamefont {Piccioli},\ and\ \citenamefont
  {Chervin}}]{le-toullec1980}%
  \BibitemOpen
  \bibfield  {author} {\bibinfo {author} {\bibfnamefont {R.}~\bibnamefont
  {Le~Toullec}}, \bibinfo {author} {\bibfnamefont {N.}~\bibnamefont
  {Piccioli}}, \ and\ \bibinfo {author} {\bibfnamefont {J.~C.}\ \bibnamefont
  {Chervin}},\ }\bibfield  {title} {\enquote {\bibinfo {title} {Optical
  properties of the band-edge exciton in {GaSe} crystals at 10 {K}},}\ }\href
  {\doibase 10.1103/PhysRevB.22.6162} {\bibfield  {journal} {\bibinfo
  {journal} {Phys. Rev. B}\ }\textbf {\bibinfo {volume} {22}},\ \bibinfo
  {pages} {6162} (\bibinfo {year} {1980})}\BibitemShut {NoStop}%
\bibitem [{\citenamefont {Gamarts}\ \emph {et~al.}(1977)\citenamefont
  {Gamarts}, \citenamefont {Ivchenko}, \citenamefont {Karaman}, \citenamefont
  {Mushinskii}, \citenamefont {Pikus}, \citenamefont {Razbirin},\ and\
  \citenamefont {Starukhin}}]{gamarts1977}%
  \BibitemOpen
  \bibfield  {author} {\bibinfo {author} {\bibfnamefont {E.~M.}\ \bibnamefont
  {Gamarts}}, \bibinfo {author} {\bibfnamefont {E.~L.}\ \bibnamefont
  {Ivchenko}}, \bibinfo {author} {\bibfnamefont {M.~I.}\ \bibnamefont
  {Karaman}}, \bibinfo {author} {\bibfnamefont {V.~P.}\ \bibnamefont
  {Mushinskii}}, \bibinfo {author} {\bibfnamefont {G.~E.}\ \bibnamefont
  {Pikus}}, \bibinfo {author} {\bibfnamefont {B.~S.}\ \bibnamefont {Razbirin}},
  \ and\ \bibinfo {author} {\bibfnamefont {A.~N.}\ \bibnamefont {Starukhin}},\
  }\bibfield  {title} {\enquote {\bibinfo {title} {Optical orientation and
  alignment of free excitons in {GaSe} during resonance excitation.
  experiment},}\ }\href@noop {} {\bibfield  {journal} {\bibinfo  {journal}
  {Sov. Phys. JETP}\ }\textbf {\bibinfo {volume} {46}},\ \bibinfo {pages} {590}
  (\bibinfo {year} {1977})}\BibitemShut {NoStop}%
\bibitem [{\citenamefont {Ivchenko}\ \emph {et~al.}(1977)\citenamefont
  {Ivchenko}, \citenamefont {Pikus}, \citenamefont {Razbirin},\ and\
  \citenamefont {Starukhin}}]{ivchenko1977}%
  \BibitemOpen
  \bibfield  {author} {\bibinfo {author} {\bibfnamefont {E.~L.}\ \bibnamefont
  {Ivchenko}}, \bibinfo {author} {\bibfnamefont {G.~E.}\ \bibnamefont {Pikus}},
  \bibinfo {author} {\bibfnamefont {B.~S.}\ \bibnamefont {Razbirin}}, \ and\
  \bibinfo {author} {\bibfnamefont {A.~I.}\ \bibnamefont {Starukhin}},\
  }\bibfield  {title} {\enquote {\bibinfo {title} {Optical orientation and
  alignment of free excitons in {GaSe} under resonant excitation. theory.}}\
  }\href@noop {} {\bibfield  {journal} {\bibinfo  {journal} {Sov. Phys. JETP}\
  }\textbf {\bibinfo {volume} {45}},\ \bibinfo {pages} {1172--1180} (\bibinfo
  {year} {1977})}\BibitemShut {NoStop}%
\bibitem [{\citenamefont {Tang}\ \emph {et~al.}(2014)\citenamefont {Tang},
  \citenamefont {Xie}, \citenamefont {Mandal}, \citenamefont {McGuire},\ and\
  \citenamefont {Lai}}]{tang2014}%
  \BibitemOpen
  \bibfield  {author} {\bibinfo {author} {\bibfnamefont {Y.}~\bibnamefont
  {Tang}}, \bibinfo {author} {\bibfnamefont {W.}~\bibnamefont {Xie}}, \bibinfo
  {author} {\bibfnamefont {K.~C.}\ \bibnamefont {Mandal}}, \bibinfo {author}
  {\bibfnamefont {J.~A.}\ \bibnamefont {McGuire}}, \ and\ \bibinfo {author}
  {\bibfnamefont {C.~W.}\ \bibnamefont {Lai}},\ }\bibfield  {title} {\enquote
  {\bibinfo {title} {Near unity optical spin polarization in {GaSe}
  nanoslabs},}\ }\href {http://arxiv.org/abs/1410.5523} {\bibfield  {journal}
  {\bibinfo  {journal} {arXiv:1410.5523}\ } (\bibinfo {year}
  {2014})}\BibitemShut {NoStop}%
\bibitem [{\citenamefont {Meier}\ and\ \citenamefont
  {Zakharchenya}(1984)}]{meier1984}%
  \BibitemOpen
  \bibfield  {author} {\bibinfo {author} {\bibfnamefont {F.}~\bibnamefont
  {Meier}}\ and\ \bibinfo {author} {\bibfnamefont {B.~P.}\ \bibnamefont
  {Zakharchenya}},\ }\bibfield  {title} {\enquote {\bibinfo {title} {Optical
  orientation},}\ \ }(\bibinfo  {publisher} {Elsevier},\ \bibinfo {year}
  {1984})\BibitemShut {NoStop}%
\bibitem [{\citenamefont {Hu}\ \emph {et~al.}(2001)\citenamefont {Hu},
  \citenamefont {Li}, \citenamefont {Yang}, \citenamefont {Manna},
  \citenamefont {Wang},\ and\ \citenamefont {Alivisatos}}]{hu2001}%
  \BibitemOpen
  \bibfield  {author} {\bibinfo {author} {\bibfnamefont {J.}~\bibnamefont
  {Hu}}, \bibinfo {author} {\bibfnamefont {L.-S.}\ \bibnamefont {Li}}, \bibinfo
  {author} {\bibfnamefont {W.}~\bibnamefont {Yang}}, \bibinfo {author}
  {\bibfnamefont {L.}~\bibnamefont {Manna}}, \bibinfo {author} {\bibfnamefont
  {L.-W.}\ \bibnamefont {Wang}}, \ and\ \bibinfo {author} {\bibfnamefont
  {A.~P.}\ \bibnamefont {Alivisatos}},\ }\bibfield  {title} {\enquote {\bibinfo
  {title} {Linearly polarized emission from colloidal semiconductor quantum
  rods},}\ }\href {\doibase 10.1126/science.1060810} {\bibfield  {journal}
  {\bibinfo  {journal} {Science}\ }\textbf {\bibinfo {volume} {292}},\ \bibinfo
  {pages} {2060--2063} (\bibinfo {year} {2001})}\BibitemShut {NoStop}%
\bibitem [{\citenamefont {Wang}\ \emph {et~al.}(2001)\citenamefont {Wang},
  \citenamefont {Gudiksen}, \citenamefont {Duan}, \citenamefont {Cui},\ and\
  \citenamefont {Lieber}}]{wang2001b}%
  \BibitemOpen
  \bibfield  {author} {\bibinfo {author} {\bibfnamefont {J.}~\bibnamefont
  {Wang}}, \bibinfo {author} {\bibfnamefont {M.~S.}\ \bibnamefont {Gudiksen}},
  \bibinfo {author} {\bibfnamefont {X.}~\bibnamefont {Duan}}, \bibinfo {author}
  {\bibfnamefont {Y.}~\bibnamefont {Cui}}, \ and\ \bibinfo {author}
  {\bibfnamefont {C.~M.}\ \bibnamefont {Lieber}},\ }\bibfield  {title}
  {\enquote {\bibinfo {title} {Highly polarized photoluminescence and
  photodetection from single indium phosphide nanowires},}\ }\href {\doibase
  10.1126/science.1062340} {\bibfield  {journal} {\bibinfo  {journal}
  {Science}\ }\textbf {\bibinfo {volume} {293}},\ \bibinfo {pages} {1455--1457}
  (\bibinfo {year} {2001})}\BibitemShut {NoStop}%
\bibitem [{\citenamefont {Johnson}\ \emph {et~al.}(2003)\citenamefont
  {Johnson}, \citenamefont {Yan}, \citenamefont {Yang},\ and\ \citenamefont
  {Saykally}}]{johnson2003}%
  \BibitemOpen
  \bibfield  {author} {\bibinfo {author} {\bibfnamefont {J.~C.}\ \bibnamefont
  {Johnson}}, \bibinfo {author} {\bibfnamefont {H.}~\bibnamefont {Yan}},
  \bibinfo {author} {\bibfnamefont {P.}~\bibnamefont {Yang}}, \ and\ \bibinfo
  {author} {\bibfnamefont {R.~J.}\ \bibnamefont {Saykally}},\ }\bibfield
  {title} {\enquote {\bibinfo {title} {Optical cavity effects in {ZnO} nanowire
  lasers and waveguides},}\ }\href {\doibase 10.1021/jp034482n} {\bibfield
  {journal} {\bibinfo  {journal} {J. Phys. Chem. B}\ }\textbf {\bibinfo
  {volume} {107}},\ \bibinfo {pages} {8816--8828} (\bibinfo {year}
  {2003})}\BibitemShut {NoStop}%
\bibitem [{\citenamefont {Takagahara}(1993)}]{takagahara1993}%
  \BibitemOpen
  \bibfield  {author} {\bibinfo {author} {\bibfnamefont {T.}~\bibnamefont
  {Takagahara}},\ }\bibfield  {title} {\enquote {\bibinfo {title} {Effects of
  dielectric confinement and electron-hole exchange interaction on excitonic
  states in semiconductor quantum dots},}\ }\href {\doibase
  10.1103/PhysRevB.47.4569} {\bibfield  {journal} {\bibinfo  {journal} {Phys.
  Rev. B}\ }\textbf {\bibinfo {volume} {47}},\ \bibinfo {pages} {4569--4585}
  (\bibinfo {year} {1993})}\BibitemShut {NoStop}%
\bibitem [{\citenamefont {Gammon}\ \emph {et~al.}(1996)\citenamefont {Gammon},
  \citenamefont {Snow}, \citenamefont {Shanabrook}, \citenamefont {Katzer},\
  and\ \citenamefont {Park}}]{gammon1996}%
  \BibitemOpen
  \bibfield  {author} {\bibinfo {author} {\bibfnamefont {D.}~\bibnamefont
  {Gammon}}, \bibinfo {author} {\bibfnamefont {E.~S.}\ \bibnamefont {Snow}},
  \bibinfo {author} {\bibfnamefont {B.~V.}\ \bibnamefont {Shanabrook}},
  \bibinfo {author} {\bibfnamefont {D.~S.}\ \bibnamefont {Katzer}}, \ and\
  \bibinfo {author} {\bibfnamefont {D.}~\bibnamefont {Park}},\ }\bibfield
  {title} {\enquote {\bibinfo {title} {Homogeneous linewidths in the optical
  spectrum of a single gallium arsenide quantum dot},}\ }\href {\doibase
  10.1126/science.273.5271.87} {\bibfield  {journal} {\bibinfo  {journal}
  {Science}\ }\textbf {\bibinfo {volume} {273}},\ \bibinfo {pages} {87--90}
  (\bibinfo {year} {1996})}\BibitemShut {NoStop}%
\bibitem [{\citenamefont {Dzhioev}\ \emph {et~al.}(1998)\citenamefont
  {Dzhioev}, \citenamefont {Zakharchenya}, \citenamefont {Ivchenko},
  \citenamefont {Korenev}, \citenamefont {Kusraev}, \citenamefont {Ledentsov},
  \citenamefont {Ustinov}, \citenamefont {Zhukov},\ and\ \citenamefont
  {Tsatsul'nikov}}]{dzhioev1998}%
  \BibitemOpen
  \bibfield  {author} {\bibinfo {author} {\bibfnamefont {R.~I.}\ \bibnamefont
  {Dzhioev}}, \bibinfo {author} {\bibfnamefont {B.~P.}\ \bibnamefont
  {Zakharchenya}}, \bibinfo {author} {\bibfnamefont {E.~L.}\ \bibnamefont
  {Ivchenko}}, \bibinfo {author} {\bibfnamefont {V.~L.}\ \bibnamefont
  {Korenev}}, \bibinfo {author} {\bibfnamefont {Y.~G.}\ \bibnamefont
  {Kusraev}}, \bibinfo {author} {\bibfnamefont {N.~N.}\ \bibnamefont
  {Ledentsov}}, \bibinfo {author} {\bibfnamefont {V.~M.}\ \bibnamefont
  {Ustinov}}, \bibinfo {author} {\bibfnamefont {A.~E.}\ \bibnamefont {Zhukov}},
  \ and\ \bibinfo {author} {\bibfnamefont {A.~F.}\ \bibnamefont
  {Tsatsul'nikov}},\ }\bibfield  {title} {\enquote {\bibinfo {title} {Optical
  orientation and alignment of excitons in quantum dots},}\ }\href {\doibase
  10.1134/1.1130397} {\bibfield  {journal} {\bibinfo  {journal} {Phys. Solid
  State}\ }\textbf {\bibinfo {volume} {40}},\ \bibinfo {pages} {790--793}
  (\bibinfo {year} {1998})}\BibitemShut {NoStop}%
\bibitem [{\citenamefont {Paillard}\ \emph {et~al.}(2001)\citenamefont
  {Paillard}, \citenamefont {Marie}, \citenamefont {Renucci}, \citenamefont
  {Amand}, \citenamefont {Jbeli},\ and\ \citenamefont {Gerard}}]{paillard2001}%
  \BibitemOpen
  \bibfield  {author} {\bibinfo {author} {\bibfnamefont {M.}~\bibnamefont
  {Paillard}}, \bibinfo {author} {\bibfnamefont {X.}~\bibnamefont {Marie}},
  \bibinfo {author} {\bibfnamefont {P.}~\bibnamefont {Renucci}}, \bibinfo
  {author} {\bibfnamefont {T.}~\bibnamefont {Amand}}, \bibinfo {author}
  {\bibfnamefont {A.}~\bibnamefont {Jbeli}}, \ and\ \bibinfo {author}
  {\bibfnamefont {J.~M.}\ \bibnamefont {Gerard}},\ }\bibfield  {title}
  {\enquote {\bibinfo {title} {Spin relaxation quenching in semiconductor
  quantum dots},}\ }\href {\doibase 10.1103/PhysRevLett.86.1634} {\bibfield
  {journal} {\bibinfo  {journal} {Phys. Rev. Lett.}\ }\textbf {\bibinfo
  {volume} {86}},\ \bibinfo {pages} {1634--1637} (\bibinfo {year}
  {2001})}\BibitemShut {NoStop}%
\bibitem [{\citenamefont {Schuller}\ \emph {et~al.}(2013)\citenamefont
  {Schuller}, \citenamefont {Karaveli}, \citenamefont {Schiros}, \citenamefont
  {He}, \citenamefont {Yang}, \citenamefont {Kymissis}, \citenamefont {Shan},\
  and\ \citenamefont {Zia}}]{schuller2013}%
  \BibitemOpen
  \bibfield  {author} {\bibinfo {author} {\bibfnamefont {J.~A.}\ \bibnamefont
  {Schuller}}, \bibinfo {author} {\bibfnamefont {S.}~\bibnamefont {Karaveli}},
  \bibinfo {author} {\bibfnamefont {T.}~\bibnamefont {Schiros}}, \bibinfo
  {author} {\bibfnamefont {K.}~\bibnamefont {He}}, \bibinfo {author}
  {\bibfnamefont {S.}~\bibnamefont {Yang}}, \bibinfo {author} {\bibfnamefont
  {I.}~\bibnamefont {Kymissis}}, \bibinfo {author} {\bibfnamefont
  {J.}~\bibnamefont {Shan}}, \ and\ \bibinfo {author} {\bibfnamefont
  {R.}~\bibnamefont {Zia}},\ }\bibfield  {title} {\enquote {\bibinfo {title}
  {Orientation of luminescent excitons in layered nanomaterials},}\ }\href
  {\doibase 10.1038/nnano.2013.20} {\bibfield  {journal} {\bibinfo  {journal}
  {Nature Nanotechno.}\ }\textbf {\bibinfo {volume} {8}},\ \bibinfo {pages}
  {271--276} (\bibinfo {year} {2013})}\BibitemShut {NoStop}%
\bibitem [{\citenamefont {Jones}\ \emph {et~al.}(2013)\citenamefont {Jones},
  \citenamefont {Yu}, \citenamefont {Ghimire}, \citenamefont {Wu},
  \citenamefont {Aivazian}, \citenamefont {Ross}, \citenamefont {Zhao},
  \citenamefont {Yan}, \citenamefont {Mandrus}, \citenamefont {Xiao},
  \citenamefont {Yao},\ and\ \citenamefont {Xu}}]{jones2013}%
  \BibitemOpen
  \bibfield  {author} {\bibinfo {author} {\bibfnamefont {A.~M.}\ \bibnamefont
  {Jones}}, \bibinfo {author} {\bibfnamefont {H.}~\bibnamefont {Yu}}, \bibinfo
  {author} {\bibfnamefont {N.~J.}\ \bibnamefont {Ghimire}}, \bibinfo {author}
  {\bibfnamefont {S.}~\bibnamefont {Wu}}, \bibinfo {author} {\bibfnamefont
  {G.}~\bibnamefont {Aivazian}}, \bibinfo {author} {\bibfnamefont {J.~S.}\
  \bibnamefont {Ross}}, \bibinfo {author} {\bibfnamefont {B.}~\bibnamefont
  {Zhao}}, \bibinfo {author} {\bibfnamefont {J.}~\bibnamefont {Yan}}, \bibinfo
  {author} {\bibfnamefont {D.~G.}\ \bibnamefont {Mandrus}}, \bibinfo {author}
  {\bibfnamefont {D.}~\bibnamefont {Xiao}}, \bibinfo {author} {\bibfnamefont
  {W.}~\bibnamefont {Yao}}, \ and\ \bibinfo {author} {\bibfnamefont
  {X.}~\bibnamefont {Xu}},\ }\bibfield  {title} {\enquote {\bibinfo {title}
  {Optical generation of excitonic valley coherence in monolayer {WSe$_2$}},}\
  }\href {\doibase 10.1038/nnano.2013.151} {\bibfield  {journal} {\bibinfo
  {journal} {Nature Nanotechno.}\ }\textbf {\bibinfo {volume} {8}},\ \bibinfo
  {pages} {634} (\bibinfo {year} {2013})}\BibitemShut {NoStop}%
\bibitem [{\citenamefont {Jones}\ \emph {et~al.}(2014)\citenamefont {Jones},
  \citenamefont {Yu}, \citenamefont {Ross}, \citenamefont {Klement},
  \citenamefont {Ghimire}, \citenamefont {Yan}, \citenamefont {Mandrus},
  \citenamefont {Yao},\ and\ \citenamefont {Xu}}]{jones2014}%
  \BibitemOpen
  \bibfield  {author} {\bibinfo {author} {\bibfnamefont {A.~M.}\ \bibnamefont
  {Jones}}, \bibinfo {author} {\bibfnamefont {H.}~\bibnamefont {Yu}}, \bibinfo
  {author} {\bibfnamefont {J.~S.}\ \bibnamefont {Ross}}, \bibinfo {author}
  {\bibfnamefont {P.}~\bibnamefont {Klement}}, \bibinfo {author} {\bibfnamefont
  {N.~J.}\ \bibnamefont {Ghimire}}, \bibinfo {author} {\bibfnamefont
  {J.}~\bibnamefont {Yan}}, \bibinfo {author} {\bibfnamefont {D.~G.}\
  \bibnamefont {Mandrus}}, \bibinfo {author} {\bibfnamefont {W.}~\bibnamefont
  {Yao}}, \ and\ \bibinfo {author} {\bibfnamefont {X.}~\bibnamefont {Xu}},\
  }\bibfield  {title} {\enquote {\bibinfo {title} {Spin--layer locking effects
  in optical orientation of exciton spin in bilayer {WSe$_2$}},}\ }\href
  {\doibase 10.1038/nphys2848} {\bibfield  {journal} {\bibinfo  {journal}
  {Nature Phys.}\ }\textbf {\bibinfo {volume} {10}},\ \bibinfo {pages}
  {130--134} (\bibinfo {year} {2014})}\BibitemShut {NoStop}%
\bibitem [{\citenamefont {Fernelius}(1994)}]{fernelius1994}%
  \BibitemOpen
  \bibfield  {author} {\bibinfo {author} {\bibfnamefont {N.~C.}\ \bibnamefont
  {Fernelius}},\ }\bibfield  {title} {\enquote {\bibinfo {title} {Properties of
  gallium selenide single crystal},}\ }\href {\doibase
  10.1016/0960-8974(94)90010-8} {\bibfield  {journal} {\bibinfo  {journal}
  {Prog. Crystal Growth and Charact. Mater.}\ }\textbf {\bibinfo {volume}
  {28}},\ \bibinfo {pages} {275--353} (\bibinfo {year} {1994})}\BibitemShut
  {NoStop}%
\bibitem [{\citenamefont {Abdullaev}\ \emph {et~al.}(1972)\citenamefont
  {Abdullaev}, \citenamefont {Kulevskii}, \citenamefont {Prokhorov},
  \citenamefont {Savel'Ev}, \citenamefont {Salaev},\ and\ \citenamefont
  {Smirnov}}]{abdullaev1972}%
  \BibitemOpen
  \bibfield  {author} {\bibinfo {author} {\bibfnamefont {G.~B.}\ \bibnamefont
  {Abdullaev}}, \bibinfo {author} {\bibfnamefont {L.~A.}\ \bibnamefont
  {Kulevskii}}, \bibinfo {author} {\bibfnamefont {A.~M.}\ \bibnamefont
  {Prokhorov}}, \bibinfo {author} {\bibfnamefont {A.~D.}\ \bibnamefont
  {Savel'Ev}}, \bibinfo {author} {\bibfnamefont {E.~Y.}\ \bibnamefont
  {Salaev}}, \ and\ \bibinfo {author} {\bibfnamefont {V.~V.}\ \bibnamefont
  {Smirnov}},\ }\bibfield  {title} {\enquote {\bibinfo {title} {{GaSe}, a new
  effective material for nonlinear optics},}\ }\href@noop {} {\bibfield
  {journal} {\bibinfo  {journal} {JETP Lett.}\ }\textbf {\bibinfo {volume}
  {16}},\ \bibinfo {pages} {90--95} (\bibinfo {year} {1972})}\BibitemShut
  {NoStop}%
\bibitem [{\citenamefont {Allakhverdiev}\ \emph {et~al.}(2009)\citenamefont
  {Allakhverdiev}, \citenamefont {Yetis}, \citenamefont {Ozbek}, \citenamefont
  {Baykara},\ and\ \citenamefont {Salaev}}]{allakhverdiev2009}%
  \BibitemOpen
  \bibfield  {author} {\bibinfo {author} {\bibfnamefont {K.~R.}\ \bibnamefont
  {Allakhverdiev}}, \bibinfo {author} {\bibfnamefont {M.~O.}\ \bibnamefont
  {Yetis}}, \bibinfo {author} {\bibfnamefont {S.}~\bibnamefont {Ozbek}},
  \bibinfo {author} {\bibfnamefont {T.~K.}\ \bibnamefont {Baykara}}, \ and\
  \bibinfo {author} {\bibfnamefont {E.~Y.}\ \bibnamefont {Salaev}},\ }\bibfield
   {title} {\enquote {\bibinfo {title} {Effective nonlinear {GaSe} crystal.
  optical properties and applications},}\ }\href {\doibase
  10.1134/S1054660X09050375} {\bibfield  {journal} {\bibinfo  {journal} {Laser
  Phys.}\ }\textbf {\bibinfo {volume} {19}},\ \bibinfo {pages} {1092--1104}
  (\bibinfo {year} {2009})}\BibitemShut {NoStop}%
\bibitem [{\citenamefont {Nahory}\ \emph {et~al.}(1971)\citenamefont {Nahory},
  \citenamefont {Shaklee}, \citenamefont {Leheny},\ and\ \citenamefont
  {DeWinter}}]{nahory1971}%
  \BibitemOpen
  \bibfield  {author} {\bibinfo {author} {\bibfnamefont {R.~E.}\ \bibnamefont
  {Nahory}}, \bibinfo {author} {\bibfnamefont {K.~L.}\ \bibnamefont {Shaklee}},
  \bibinfo {author} {\bibfnamefont {R.~F.}\ \bibnamefont {Leheny}}, \ and\
  \bibinfo {author} {\bibfnamefont {J.~C.}\ \bibnamefont {DeWinter}},\
  }\bibfield  {title} {\enquote {\bibinfo {title} {Stimulated emission and the
  type of bandgap in {GaSe}},}\ }\href {\doibase 10.1016/0038-1098(71)90472-8}
  {\bibfield  {journal} {\bibinfo  {journal} {Solid State Commun.}\ }\textbf
  {\bibinfo {volume} {9}},\ \bibinfo {pages} {1107--1111} (\bibinfo {year}
  {1971})}\BibitemShut {NoStop}%
\bibitem [{\citenamefont {Ugmori}\ \emph {et~al.}(1973)\citenamefont {Ugmori},
  \citenamefont {Masuda},\ and\ \citenamefont {Namba}}]{ugmori1973}%
  \BibitemOpen
  \bibfield  {author} {\bibinfo {author} {\bibfnamefont {T.}~\bibnamefont
  {Ugmori}}, \bibinfo {author} {\bibfnamefont {K.}~\bibnamefont {Masuda}}, \
  and\ \bibinfo {author} {\bibfnamefont {S.}~\bibnamefont {Namba}},\ }\bibfield
   {title} {\enquote {\bibinfo {title} {Spontaneous and stimulated emission in
  {GaSe} under intense excitation},}\ }\href {\doibase
  10.1016/0038-1098(73)90779-5} {\bibfield  {journal} {\bibinfo  {journal}
  {Solid State Commun.}\ }\textbf {\bibinfo {volume} {12}},\ \bibinfo {pages}
  {389--391} (\bibinfo {year} {1973})}\BibitemShut {NoStop}%
\bibitem [{\citenamefont {Voitchovsky}\ and\ \citenamefont
  {Mercier}(1974)}]{voitchovsky1974}%
  \BibitemOpen
  \bibfield  {author} {\bibinfo {author} {\bibfnamefont {J.~P.}\ \bibnamefont
  {Voitchovsky}}\ and\ \bibinfo {author} {\bibfnamefont {A.}~\bibnamefont
  {Mercier}},\ }\bibfield  {title} {\enquote {\bibinfo {title}
  {Photoluminescence of {GaSe}},}\ }\href {\doibase 10.1007/BF02726593}
  {\bibfield  {journal} {\bibinfo  {journal} {Nuovo Cimento B}\ }\textbf
  {\bibinfo {volume} {22}},\ \bibinfo {pages} {273--292} (\bibinfo {year}
  {1974})}\BibitemShut {NoStop}%
\bibitem [{\citenamefont {Mercier}\ and\ \citenamefont
  {Voitchovsky}(1975)}]{mercier1975}%
  \BibitemOpen
  \bibfield  {author} {\bibinfo {author} {\bibfnamefont {A.}~\bibnamefont
  {Mercier}}\ and\ \bibinfo {author} {\bibfnamefont {J.~P.}\ \bibnamefont
  {Voitchovsky}},\ }\bibfield  {title} {\enquote {\bibinfo {title}
  {Exciton-exciton and exciton-carrier scattering in {GaSe}},}\ }\href
  {\doibase 10.1103/PhysRevB.11.2243} {\bibfield  {journal} {\bibinfo
  {journal} {Phys. Rev. B}\ }\textbf {\bibinfo {volume} {11}},\ \bibinfo
  {pages} {2243--2250} (\bibinfo {year} {1975})}\BibitemShut {NoStop}%
\bibitem [{\citenamefont {Moriya}\ and\ \citenamefont
  {Kushida}(1976{\natexlab{a}})}]{moriya1976}%
  \BibitemOpen
  \bibfield  {author} {\bibinfo {author} {\bibfnamefont {T.}~\bibnamefont
  {Moriya}}\ and\ \bibinfo {author} {\bibfnamefont {T.}~\bibnamefont
  {Kushida}},\ }\bibfield  {title} {\enquote {\bibinfo {title} {Luminescence
  spectra due to exciton-exditon collisions in semiconductors.: I. spontaneous
  emission spectra},}\ }\href {\doibase 10.1143/JPSJ.40.1668} {\bibfield
  {journal} {\bibinfo  {journal} {J. Phys. Soc. Jpn.}\ }\textbf {\bibinfo
  {volume} {40}},\ \bibinfo {pages} {1668--1675} (\bibinfo {year}
  {1976}{\natexlab{a}})}\BibitemShut {NoStop}%
\bibitem [{\citenamefont {Moriya}\ and\ \citenamefont
  {Kushida}(1976{\natexlab{b}})}]{moriya1976a}%
  \BibitemOpen
  \bibfield  {author} {\bibinfo {author} {\bibfnamefont {T.}~\bibnamefont
  {Moriya}}\ and\ \bibinfo {author} {\bibfnamefont {T.}~\bibnamefont
  {Kushida}},\ }\bibfield  {title} {\enquote {\bibinfo {title} {Luminescence
  spectra due to exciton-exciton collisions in semiconductors. ii. stimulated
  emission spectra},}\ }\href {\doibase 10.1143/JPSJ.40.1676} {\bibfield
  {journal} {\bibinfo  {journal} {J. Phys. Soc. Jpn.}\ }\textbf {\bibinfo
  {volume} {40}},\ \bibinfo {pages} {1676--1683} (\bibinfo {year}
  {1976}{\natexlab{b}})}\BibitemShut {NoStop}%
\bibitem [{\citenamefont {Bernier}\ \emph {et~al.}(1986)\citenamefont
  {Bernier}, \citenamefont {Jandl},\ and\ \citenamefont
  {Provencher}}]{bernier1986}%
  \BibitemOpen
  \bibfield  {author} {\bibinfo {author} {\bibfnamefont {G.}~\bibnamefont
  {Bernier}}, \bibinfo {author} {\bibfnamefont {S.}~\bibnamefont {Jandl}}, \
  and\ \bibinfo {author} {\bibfnamefont {R.}~\bibnamefont {Provencher}},\
  }\bibfield  {title} {\enquote {\bibinfo {title} {Spontaneous and stimulated
  photoluminescence of {GaSe} in the energy range 2.075--2.125 {eV}},}\ }\href
  {\doibase 10.1016/0022-2313(86)90052-9} {\bibfield  {journal} {\bibinfo
  {journal} {J. Lumin.}\ }\textbf {\bibinfo {volume} {35}},\ \bibinfo {pages}
  {289--300} (\bibinfo {year} {1986})}\BibitemShut {NoStop}%
\bibitem [{\citenamefont {Pavesi}\ \emph {et~al.}(1989)\citenamefont {Pavesi},
  \citenamefont {Staehli},\ and\ \citenamefont {Capozzi}}]{pavesi1989}%
  \BibitemOpen
  \bibfield  {author} {\bibinfo {author} {\bibfnamefont {L.}~\bibnamefont
  {Pavesi}}, \bibinfo {author} {\bibfnamefont {J.~L.}\ \bibnamefont {Staehli}},
  \ and\ \bibinfo {author} {\bibfnamefont {V.}~\bibnamefont {Capozzi}},\
  }\bibfield  {title} {\enquote {\bibinfo {title} {Mott transition of the
  excitons in {GaSe}},}\ }\href {\doibase 10.1103/PhysRevB.39.10982} {\bibfield
   {journal} {\bibinfo  {journal} {Phys. Rev. B}\ }\textbf {\bibinfo {volume}
  {39}},\ \bibinfo {pages} {10982--10994} (\bibinfo {year} {1989})}\BibitemShut
  {NoStop}%
\bibitem [{\citenamefont {Capozzi}\ \emph {et~al.}(1993)\citenamefont
  {Capozzi}, \citenamefont {Pavesi},\ and\ \citenamefont
  {Staehli}}]{capozzi1993}%
  \BibitemOpen
  \bibfield  {author} {\bibinfo {author} {\bibfnamefont {V.}~\bibnamefont
  {Capozzi}}, \bibinfo {author} {\bibfnamefont {L.}~\bibnamefont {Pavesi}}, \
  and\ \bibinfo {author} {\bibfnamefont {J.~L.}\ \bibnamefont {Staehli}},\
  }\bibfield  {title} {\enquote {\bibinfo {title} {Exciton-carrier scattering
  in gallium selenide},}\ }\href {\doibase 10.1103/PhysRevB.47.6340} {\bibfield
   {journal} {\bibinfo  {journal} {Phys. Rev. B.}\ }\textbf {\bibinfo {volume}
  {47}},\ \bibinfo {pages} {6340--6349} (\bibinfo {year} {1993})}\BibitemShut
  {NoStop}%
\bibitem [{\citenamefont {Mandal}\ \emph {et~al.}(2008)\citenamefont {Mandal},
  \citenamefont {Mertiri}, \citenamefont {Pabst}, \citenamefont {Roy},
  \citenamefont {Cui}, \citenamefont {Battacharya}, \citenamefont {Groza},
  \citenamefont {Burger}, \citenamefont {Conway},\ and\ \citenamefont
  {Nikolic}}]{mandal2008a}%
  \BibitemOpen
  \bibfield  {author} {\bibinfo {author} {\bibfnamefont {K.~C.}\ \bibnamefont
  {Mandal}}, \bibinfo {author} {\bibfnamefont {A.}~\bibnamefont {Mertiri}},
  \bibinfo {author} {\bibfnamefont {G.~W.}\ \bibnamefont {Pabst}}, \bibinfo
  {author} {\bibfnamefont {R.~G.}\ \bibnamefont {Roy}}, \bibinfo {author}
  {\bibfnamefont {Y.}~\bibnamefont {Cui}}, \bibinfo {author} {\bibfnamefont
  {P.}~\bibnamefont {Battacharya}}, \bibinfo {author} {\bibfnamefont
  {M.}~\bibnamefont {Groza}}, \bibinfo {author} {\bibfnamefont
  {A.}~\bibnamefont {Burger}}, \bibinfo {author} {\bibfnamefont {A.~M.}\
  \bibnamefont {Conway}}, \ and\ \bibinfo {author} {\bibfnamefont {R.~J.}\
  \bibnamefont {Nikolic}},\ }\bibfield  {title} {\enquote {\bibinfo {title}
  {Layered {III-VI} chalcogenide semiconductor crystals for radiation
  detectors},}\ }in\ \href {\doibase 10.1117/12.796235} {\emph {\bibinfo
  {booktitle} {Proc. SPIE}}},\ Vol.\ \bibinfo {volume} {7079}\ (\bibinfo {year}
  {2008})\ p.\ \bibinfo {pages} {70790O}\BibitemShut {NoStop}%
\bibitem [{\citenamefont {Marcuse}(1991)}]{marcuse1991}%
  \BibitemOpen
  \bibfield  {author} {\bibinfo {author} {\bibfnamefont {D.}~\bibnamefont
  {Marcuse}},\ }\enquote {\bibinfo {title} {The asymmetric slab waveguide},}\
  in\ \href@noop {} {\emph {\bibinfo {booktitle} {Quantum electronics --
  Principles and applications}}}\ (\bibinfo {year} {1991})\ pp.\ \bibinfo
  {pages} {1--59}\BibitemShut {NoStop}%
\bibitem [{\citenamefont {Guti{\'e}rrez}\ \emph {et~al.}(2012)\citenamefont
  {Guti{\'e}rrez}, \citenamefont {Perea-L{\'o}pez}, \citenamefont {El{\'\i}as},
  \citenamefont {Berkdemir}, \citenamefont {Wang}, \citenamefont {Ruitao},
  \citenamefont {L{\'o}pez-Ur{\'\i}as}, \citenamefont {Crespi}, \citenamefont
  {Terrones},\ and\ \citenamefont {Terrones}}]{gutierrez2012}%
  \BibitemOpen
  \bibfield  {author} {\bibinfo {author} {\bibfnamefont {H.~R.}\ \bibnamefont
  {Guti{\'e}rrez}}, \bibinfo {author} {\bibfnamefont {N.}~\bibnamefont
  {Perea-L{\'o}pez}}, \bibinfo {author} {\bibfnamefont {A.~L.}\ \bibnamefont
  {El{\'\i}as}}, \bibinfo {author} {\bibfnamefont {A.}~\bibnamefont
  {Berkdemir}}, \bibinfo {author} {\bibfnamefont {B.}~\bibnamefont {Wang}},
  \bibinfo {author} {\bibfnamefont {Lv.}\ \bibnamefont {Ruitao}}, \bibinfo
  {author} {\bibfnamefont {F.}~\bibnamefont {L{\'o}pez-Ur{\'\i}as}}, \bibinfo
  {author} {\bibfnamefont {V.~H.}\ \bibnamefont {Crespi}}, \bibinfo {author}
  {\bibfnamefont {H.}~\bibnamefont {Terrones}}, \ and\ \bibinfo {author}
  {\bibfnamefont {M.}~\bibnamefont {Terrones}},\ }\bibfield  {title} {\enquote
  {\bibinfo {title} {Extraordinary room-temperature photoluminescence in
  triangular {WS$_2$} monolayers},}\ }\href {\doibase 10.1021/nl3026357}
  {\bibfield  {journal} {\bibinfo  {journal} {Nano Lett.}\ }\textbf {\bibinfo
  {volume} {13}},\ \bibinfo {pages} {3447--3454} (\bibinfo {year}
  {2012})}\BibitemShut {NoStop}%
\end{thebibliography}
\end{document}